\documentclass[conference]{IEEEtran}
\IEEEoverridecommandlockouts

\usepackage{cite}
\usepackage{amsmath,amssymb,amsfonts}
\usepackage{algorithmic}
\usepackage{graphicx}
\usepackage{textcomp}
\usepackage{xcolor}
\usepackage{marvosym}

\usepackage{lipsum}
\usepackage{booktabs}
\usepackage{makecell}
\usepackage{multirow}
\usepackage{pifont}
\usepackage[linesnumbered,ruled,noend]{algorithm2e}
\SetKw{Continue}{continue}
\SetKw{And}{and}
\usepackage[colorlinks, linkcolor=magenta, anchorcolor=green, citecolor=blue]{hyperref}

\usepackage[top=0.7in,bottom=0.82in,left=0.65in,right=0.65in]{geometry}

\def\BibTeX{{\rm B\kern-.05em{\sc i\kern-.025em b}\kern-.08em
    T\kern-.1667em\lower.7ex\hbox{E}\kern-.125emX}}
\begin{document}

\title{\huge CIMFlow: An Integrated Framework for Systematic Design and Evaluation of Digital CIM Architectures}
\author{\IEEEauthorblockN{
Yingjie Qi\IEEEauthorrefmark{2},
Jianlei Yang\IEEEauthorrefmark{2}\IEEEauthorrefmark{4}\textsuperscript{\Letter},
Yiou Wang\IEEEauthorrefmark{2}, 
Yikun Wang\IEEEauthorrefmark{2},
Dayu Wang\IEEEauthorrefmark{2}, 
Ling Tang\IEEEauthorrefmark{2}, \\
Cenlin Duan\IEEEauthorrefmark{3},
Xiaolin He\IEEEauthorrefmark{2},
Weisheng Zhao\IEEEauthorrefmark{3}}
\IEEEauthorblockA{\IEEEauthorrefmark{2} School of Computer Science and Engineering, Beihang University}
\IEEEauthorblockA{\IEEEauthorrefmark{3} School of Integrated Circuit Science and Engineering, Beihang University}
\IEEEauthorblockA{\IEEEauthorrefmark{4} Qingdao Research Institute, Beihang University}
\thanks{This work is supported in part by the National Natural Science Foundation of China (Grant No. 62072019), the Beijing Natural Science Foundation (Grant No. L243031), and the National Key R\&D Program of China (Grant No. 2023YFB4503704 and 2024YFB4505601).}
\thanks{\textsuperscript{\Letter}Corresponding author is Jianlei Yang. Email: \href{mailto:jianlei@buaa.edu.cn}{jianlei@buaa.edu.cn}.}
\thanks{The CIMFlow framework and its source code will soon be made available at \url{https://cimflow.org/}.}
\vspace{-5pt}
}

\maketitle

\bstctlcite{IEEEexample:BSTcontrol}

\begin{abstract}
Digital Compute-in-Memory (CIM) architectures have shown great promise in Deep Neural Network (DNN) acceleration by effectively addressing the “memory wall” bottleneck.
However, the development and optimization of digital CIM accelerators are hindered by the lack of comprehensive tools that encompass both software and hardware design spaces.
Moreover, existing design and evaluation frameworks often lack support for the capacity constraints inherent in digital CIM architectures.
In this paper, we present CIMFlow, an integrated framework that provides an out-of-the-box workflow for implementing and evaluating DNN workloads on digital CIM architectures.
CIMFlow bridges the compilation and simulation infrastructures with a flexible instruction set architecture (ISA) design, and addresses the constraints of digital CIM through advanced partitioning and parallelism strategies in the compilation flow.
Our evaluation demonstrates that CIMFlow enables systematic prototyping and optimization of digital CIM architectures across diverse configurations, providing researchers and designers with an accessible platform for extensive design space exploration.
\end{abstract}

\vspace{-5pt}
\section{Introduction}

Recent advances in Deep Neural Networks (DNNs) have led to unprecedented achievements across various domains, significantly increasing the demand for efficient processing of deep learning workloads. 
However, traditional von Neumann architectures~\cite{chen2016eyeriss, yang2021s} are hitting the “memory wall” due to frequent data transmission between separate computing and memory units~\cite{gholami2024ai}. 
As AI technologies continue to transform computing, Compute-in-Memory (CIM) architectures~\cite{chi2016prime, zhao2023nand, duan2023ddc} have emerged as a promising solution for next-generation DNN accelerators, aiming to eliminate this bottleneck by integrating computation logic within memory arrays.

Mainstream CIM designs can be broadly categorized into analog~\cite{shafiee2016isaac, ankit2019puma, su202116} and digital~\cite{chih202116, yan20221, duan2024towards} approaches.
While analog CIM relies on current/voltage summation for computations, digital CIM embeds digital logic units directly into SRAM arrays.
By avoiding analog-to-digital (ADC) and digital-to-analog (DAC) conversion overhead and non-ideality issues inherent in analog computations, digital CIM demonstrates both robust computation and enhanced parallelism, thus exhibiting great potential in DNN acceleration.

However, fully realizing the benefits of digital CIM across various DNN workloads is fraught with challenges.
Current CIM architectures, like many domain-specific accelerators, are typically optimized for specific applications, limiting their adaptability to emerging workloads.
As the field of AI continues to evolve, designers face a constantly expanding design space, comprising diverse DNN structures, dataflow mapping strategies, and various design choices in CIM accelerators.
Navigating this complex landscape requires versatile tools capable of accommodating a wide range of software and hardware configurations.

\begin{figure}[t]
    \centering 

    \resizebox{\linewidth}{!}{
    \begin{tabular}{ccccc}
    \toprule
    & \textbf{NeuroSim [15]} & \textbf{MNSIM [16]} & \textbf{CIM-MLC [19]} & \textbf{CIMFlow} \\ 
    \midrule
    \textbf{Design Level} &
      \begin{tabular}[c]{@{}c@{}}System-level\\ simulation\end{tabular} &
      \begin{tabular}[c]{@{}c@{}}System-level\\ simulation\end{tabular} &
      \begin{tabular}[c]{@{}c@{}}Compilation\\ framework\end{tabular} &
      \begin{tabular}[c]{@{}c@{}}Integrated\\ framework\end{tabular} \\ 
    \midrule
    \begin{tabular}[c]{@{}c@{}}\textbf{SW/HW}\\ \textbf{Co-design}\end{tabular} &
      \begin{tabular}[c]{@{}c@{}}Tightly\\ coupled\end{tabular} &
      \begin{tabular}[c]{@{}c@{}}Tightly\\ coupled\end{tabular} &
      - &
      \begin{tabular}[c]{@{}c@{}}Modularized\end{tabular} \\
    \midrule
    \textbf{Datapath}    & Fixed             & Configurable   & High-level       & Extensible       \\ 
    \midrule
    \textbf{DNN Support} & Limited           & Limited        & Flexible         & Flexible         \\ 
    \midrule
    \textbf{ISA Support} &
      \begin{tabular}[c]{@{}c@{}}Not\\ supported\end{tabular} &
      \begin{tabular}[c]{@{}c@{}}Not\\ supported\end{tabular} &
      \begin{tabular}[c]{@{}c@{}}Partially \\ supported\end{tabular} &
      \begin{tabular}[c]{@{}c@{}}ISA-based\\ framework\end{tabular} \\
    \bottomrule
    \end{tabular}
    }

    \vspace{5pt}

    \includegraphics[width=\linewidth]{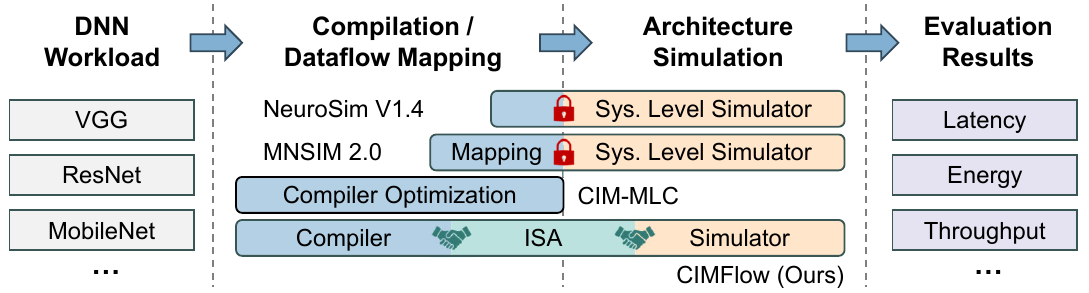}

    \caption{Comparing CIMFlow with recent design and evaluation frameworks for CIM architectures.} 
    \label{fig:relatedworks}
    \vspace{-8pt}
\end{figure}

While various software tools have been developed to facilitate CIM designs, existing approaches have notable limitations.
On the one hand, many tools tend to focus predominantly on specific aspects of the design flow, such as hardware simulation~\cite{zheng2022pimulator, liu2022simulation, lee2024neurosim, zhu2023mnsim} or dataflow compilation~\cite{siemieniuk2021occ, sun2023pimcomp, qu2024cim}, lacking the holistic view necessary for effective design space exploration.
On the other hand, most of these tools are primarily designed for analog CIM, and are only later adapted to support digital implementations, often overlooking crucial characteristics of digital CIM architectures.

In particular, SRAM-based digital CIM faces inherent capacity constraints due to its lower integration density compared to DRAM or emerging NVM solutions~\cite{jhang2021challenges}.
In practical implementations, these density limitations often result in insufficient on-chip memory capacity for modern DNN models.
This constraint could potentially negate the benefits of in-memory computing, as it necessitates frequent data movement and restricts opportunities for inter-layer pipeline optimization.
Consequently, there is a pressing need for an integrated framework specifically tailored to the nuances of digital CIM, enabling comprehensive evaluation and rapid prototyping across diverse configurations.

To address these challenges, we introduce CIMFlow, an integrated framework for systematic design and evaluation of digital CIM architectures.
As shown in Fig.~\ref{fig:relatedworks}, our framework integrates a highly extensible Instruction Set Architecture (ISA), a digital CIM-oriented compiler, and an efficient cycle-accurate simulator, enabling comprehensive and flexible exploration of diverse software and hardware configurations.
The main contributions can be summarized as follows:
\begin{itemize}
    \item We propose CIMFlow, an integrated framework that provides an out-of-the-box workflow for implementing and evaluating DNN workloads on digital CIM architectures. The framework enables systematic design space exploration through seamless integration of compilation and simulation infrastructures.
    \item We develop a CIM-specific ISA design that employs a hierarchical hardware abstraction, bridging compilation and simulation while providing flexible support for various architectural configurations.
    \item We implement an advanced compilation flow built on the MLIR infrastructure, addressing the capacity limitations in digital CIM architectures through innovative partitioning and parallelism strategies.
\end{itemize}
\section{Background and Motivation}

\subsection{Digital CIM Preliminaries}

CIM architectures represent a paradigm shift from traditional von Neumann computing by enabling computation directly within memory arrays.
In essence, CIM architectures perform matrix-vector multiplication (MVM) through parallelized in-situ Multiply-Accumulate (MAC) operations within memory arrays, significantly reducing the data movement bottleneck that plagues conventional architectures.
These MAC operations are typically carried out in a bit-serial fashion, where multiplications are decomposed into a series of bit-wise computations, followed by shift and accumulation operations.

A digital CIM macro comprises two key components: (1) a modified SRAM array that stores weight data and enables in-situ computations, and (2) peripheral circuits that orchestrate row-wise Boolean operations and accumulation\cite{chen2023autodcim}.
Unlike analog CIM approaches, which face significant limitations due to area and power overhead from ADC and DAC converters, digital CIM enables simultaneous activation of the entire memory array.
This enhanced parallelism, combined with the inherent robustness of digital computation, makes digital CIM particularly well-suited for accelerating deep learning workloads that exhibit high computational intensity and parallel structure.
While researchers have proposed various digital CIM accelerators to harness these benefits~\cite{tu2022redcim, desoli202316}, scaling these solutions to meet the diverse computational demands of modern AI models remains a significant challenge.

\subsection{Related Works and Our Motivation}

The growing complexity of deep learning workloads and CIM architectures has created a critical need for tools and frameworks to bridge the gap between hardware and software.
Existing frameworks broadly fall into two categories: modeling/simulation frameworks that evaluate architectural decisions through detailed performance analysis~\cite{zheng2022pimulator, liu2022simulation, lee2024neurosim, zhu2023mnsim}, and compilation frameworks that enable efficient mapping of DNN workloads to CIM hardware~\cite{siemieniuk2021occ, sun2023pimcomp, qu2024cim}.
While modeling frameworks help hardware designers optimize latency, power, and area trade-offs, compilation frameworks provide support for managing the hardware-software interface and generating optimized dataflows for CIM architectures.

However, most existing frameworks have focused primarily on analog CIM architectures, given their prominence in current research.
As illustrated in Fig.~\ref{fig:relatedworks}, recent simulation frameworks have begun extending support towards digital CIM implementations.
NeuroSim V1.4~\cite{lee2024neurosim} extends its system-level simulation capabilities to support advanced digital CIM technology nodes, while MNSIM 2.0~\cite{zhu2023mnsim} proposes a unified memory array model for both analog and digital CIM architectures.
Although these tools provide valuable insights into architectural trade-offs, they often employ tightly-coupled co-design with fixed assumptions about DNN structures and datapath organizations, limiting their flexibility and applicability for comprehensive digital CIM exploration.
On the compilation front, while traditional deep learning compilation infrastructures offer general-purpose solutions~\cite{lattner2021mlir}, frameworks such as CIM-MLC~\cite{qu2024cim} introduce optimization strategies specifically designed for analog CIM. 
However, there remains a notable lack of dedicated compilation support for digital CIM architectures.

The above observations highlight three key challenges in digital CIM development.
\ding{182} \underline{\textbf{Integration Gap:}} The lack of comprehensive integration between different design stages necessitates an integrated framework that combines ISA definition, compilation, and simulation.
\ding{183} \underline{\textbf{Limited Flexibility:}} Fixed architectural assumptions in current tools might constrain design space exploration, calling for extensibility in CIM design and evaluation frameworks.
\ding{184} \underline{\textbf{Resource Constraints:}} The limited SRAM array capacity in digital CIM requires sophisticated dataflow management and parallelism strategies that existing works do not adequately address.
To tackle these challenges, we propose CIMFlow, an integrated framework that enables flexible design space exploration through extensible ISA support and advanced compilation optimization techniques.
By providing comprehensive support across the design stack, CIMFlow facilitates structured development and evaluation of digital CIM architectures while maintaining the flexibility to accommodate future innovations.

\begin{figure*}[t]
    \centering
    \includegraphics[width=0.95\linewidth]{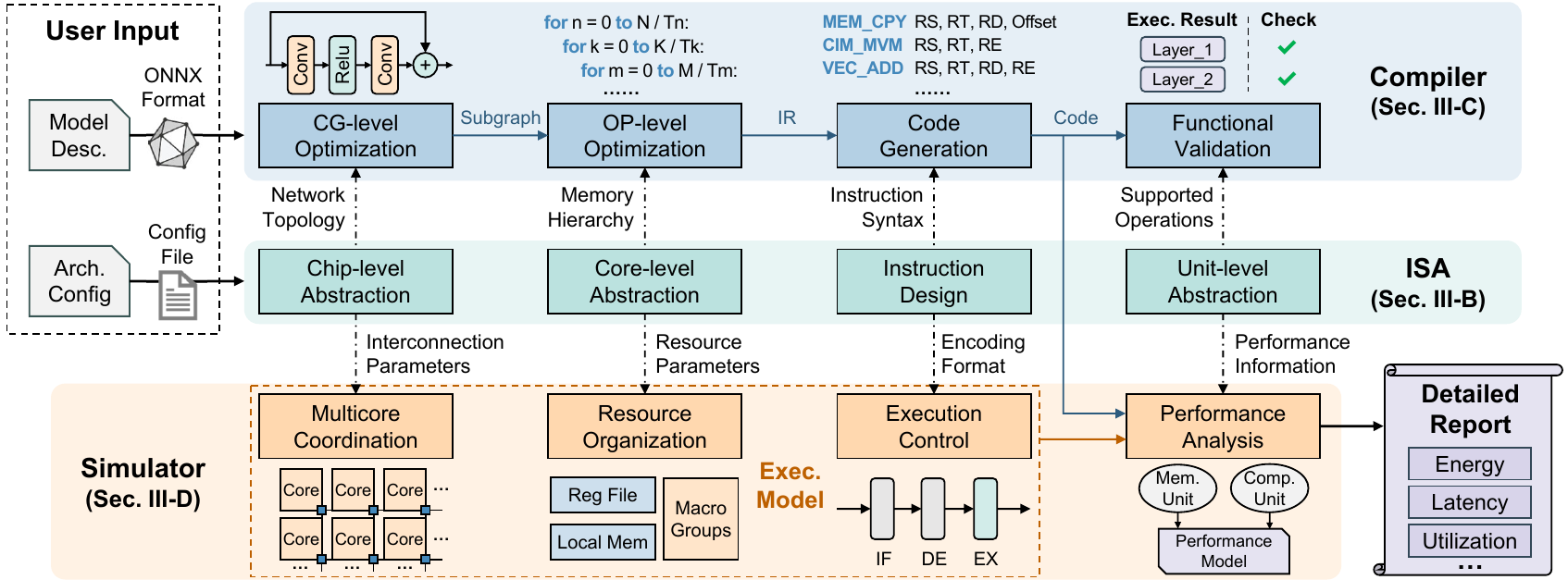}
    \caption{Overview of the CIMFlow framework.}
    \label{fig:overview}
    \vspace{-5pt}
\end{figure*}

\section{CIMFlow Framework}

\subsection{Framework Overview}

The design of CIMFlow is guided by two core principles: \textbf{integration} for seamless exploration, and \textbf{extensibility} through hierarchical abstractions.
CIMFlow unifies the entire workflow from high-level DNN model description to detailed performance analysis, providing researchers with a comprehensive view of DNN workload execution on digital CIM architectures.
Built with a modular design, CIMFlow readily adapts to emerging advances in CIM technology while abstracting implementation complexities through its streamlined interface, enabling designers and researchers to perform systematic evaluation and exploration.

As illustrated in Fig.~\ref{fig:overview}, CIMFlow offers a cohesive workflow through three main components: a flexible ISA with hierarchical hardware abstraction (Sec.~\ref{sec:isa}), a compiler featuring multilevel optimization techniques (Sec.~\ref{sec:compiler}), and a cycle-accurate simulator delivering detailed performance insights (Sec.~\ref{sec:simulator}).
The workflow begins with a DNN model description in ONNX format, complemented by an architecture configuration file that specifies the target CIM hardware parameters.
The compiler first performs computational graph (CG) level optimizations for workload distribution and data management, followed by operator (OP) level optimizations to further maximize hardware efficiency.
These optimizations are guided by the hardware specifications defined through the hierarchical hardware abstraction interface in ISA, spanning from chip-level interconnection to unit-level execution details. 
The generated code is validated by the compiler and then fed into the cycle-accurate simulator, which models the execution across multiple cores while tracking resource utilization and performance metrics. 
This process produces a detailed report covering energy consumption, latency, and hardware utilization, enabling comprehensive evaluation of different architecture designs.

\subsection{ISA Design}\label{sec:isa}

\begin{figure}[t]
    \centering
    \includegraphics[width=\linewidth]{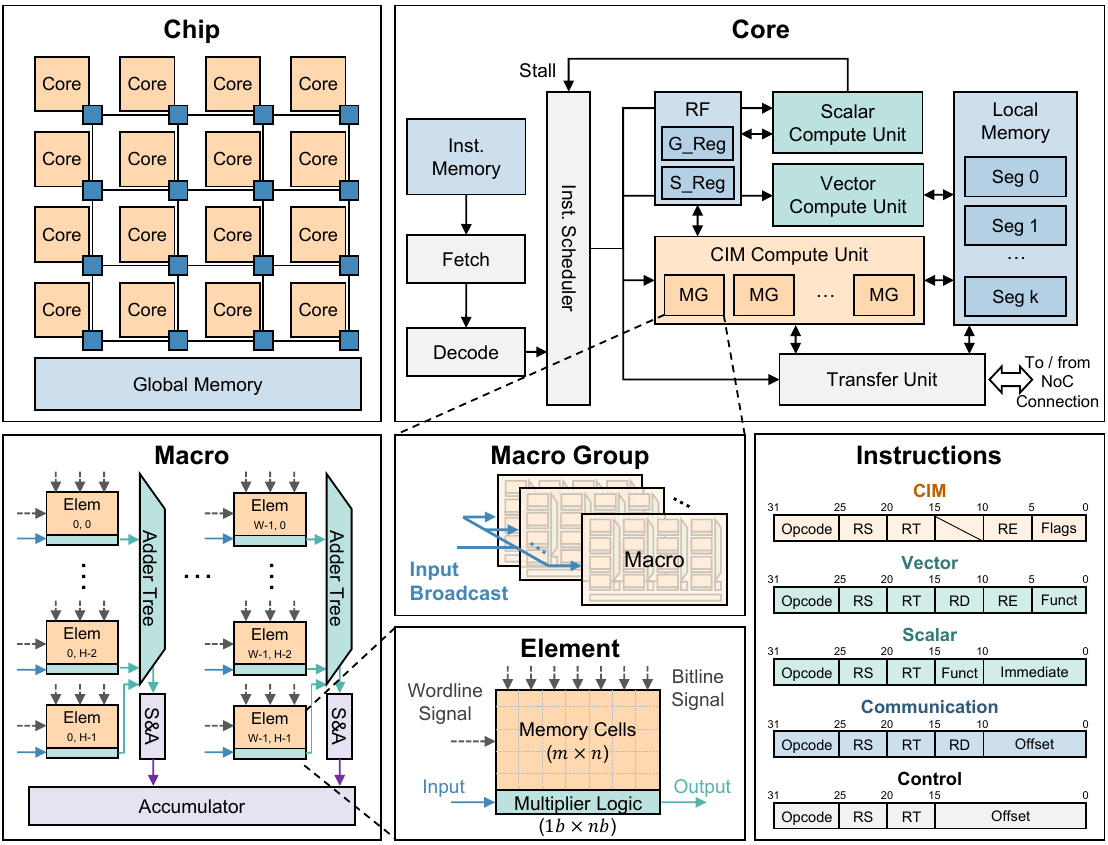}
    \caption{Hardware abstraction and instruction design in the CIMFlow ISA.}
    \label{fig:isa}
\end{figure}

As depicted in Fig.~\ref{fig:isa}, the CIMFlow ISA implements a three-level hardware abstraction hierarchy complemented by a flexible instruction set design.
Each abstraction level interfaces with the corresponding stages in the compilation and simulation infrastructure, providing architectural specifications that guide both compilation optimization and simulation execution.

\textbf{Hardware Abstraction.}
At chip level, the architecture consists of multiple cores interconnected through a Network-on-Chip (NoC) structure, facilitating synchronous inter-core communication and global memory access.
This organization enables scalable workload distribution and flexible inter-core pipelining, with each core functioning as a basic unit of program execution with its own instruction control flow.

The core-level abstraction defines the organization of hardware resources, encompassing  instruction memory, various compute units, register files (RFs), and local memory. 
To facilitate efficient memory management and architectural extensibility, CIMFlow implements a unified address space across both global and local memories. 
In addition, the local memory is divided into segments to efficiently handle the input and output of DNN layers.
The register file consists of general-purpose registers (G\_Reg) for instruction-level access and special-purpose registers (S\_Reg) for operation-specific functions.

At unit level, the CIM compute unit incorporates multiple macro groups (MGs) that support weight duplication and flexible spatial mapping strategies. 
Within each MG, weights are typically organized along the output channel, enabling efficient input data broadcast across macros for parallel in-memory MVM operations.
The vector compute unit handles auxiliary DNN operations such as activation, pooling, and quantization.
The scalar compute unit executes control flow operations through scalar arithmetic computations.

\textbf{Instruction Design.}
To support efficient execution across the hardware hierarchy, CIMFlow implements a unified 32-bit instruction format with specialized variations for different operation types. 
Instructions are categorized into compute, communication, and control flow instructions, with compute instructions further specialized for CIM, vector, and scalar compute units. 
Each instruction contains a 6-bit operation specifier (opcode) and multiple 5-bit operand fields.
Certain instruction types may also include supplementary fields, such as a 6-bit functionality specifier, execution flags, or immediate values of 10 or 16 bits.
The instruction format supports up to four operands depending on the operation type, providing flexibility for complex operations while maintaining encoding efficiency.
The instruction set is designed for extensibility through incorporating a customized instruction description template, which enables seamless integration of new operations into the framework when provided with their associated performance parameters.

\subsection{Compilation Flow}\label{sec:compiler}

\begin{figure}[t]
    \centering
    \includegraphics[width=\linewidth]{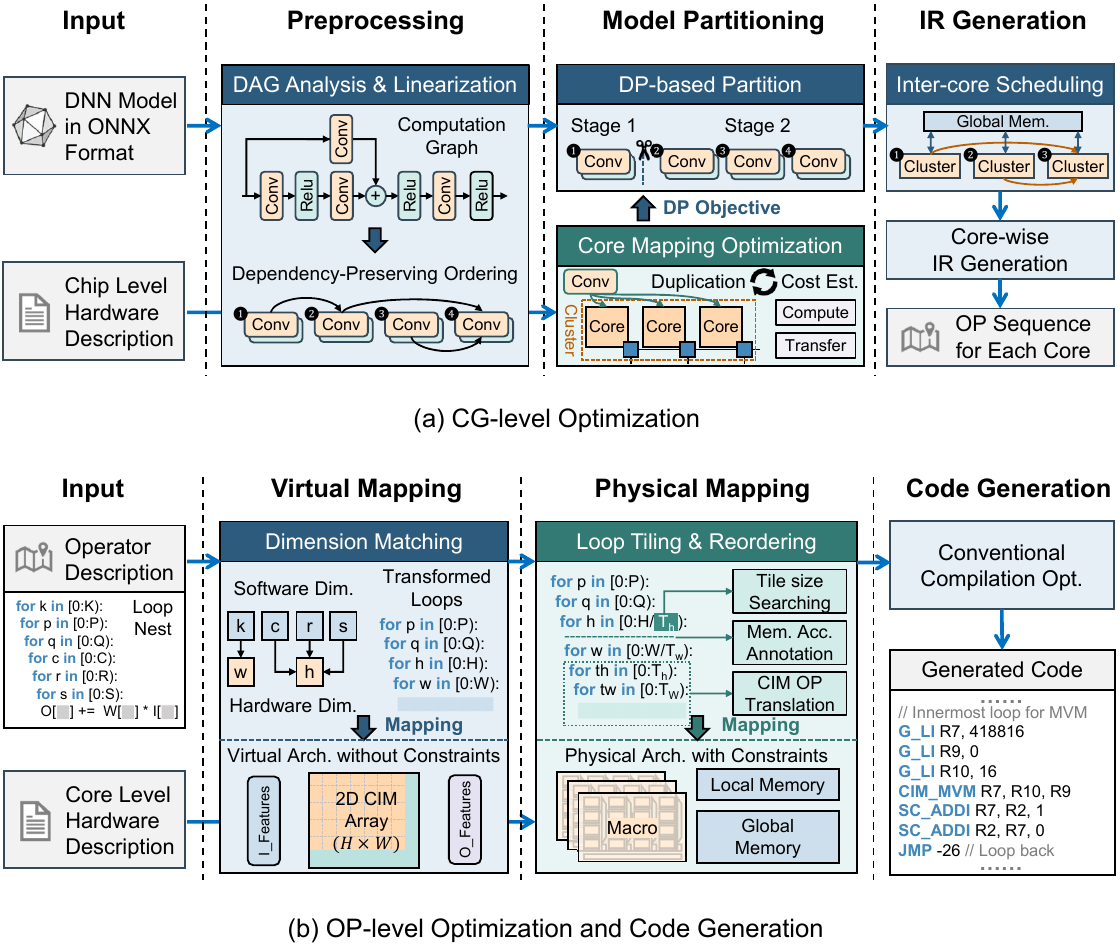}
    \caption{Compilation flow and mapping optimization strategies in CIMFlow.}
    \label{fig:compiler}
\end{figure}

The CIMFlow compiler bridges the semantic gap between high-level DNN models and low-level CIM operations through a two-level optimization strategy, as illustrated in Fig.~\ref{fig:compiler}.
Starting with an ONNX model, the compiler first performs CG-level optimizations to partition and schedule workloads across multiple cores, effectively addressing the limited capacity issue. 
This is followed by OP-level optimizations built upon the MLIR infrastructure~\cite{lattner2021mlir}, which translates DNN operations into efficient CIM instruction sequences while taking into account the underlying hardware constraints.

\begin{algorithm}[t]
\SetAlgoLined
\KwIn{Preprocessed computation graph $G=(V,E)$, Hardware resources $R$}
\KwOut{Optimal partitioning and mapping solution $S$}
$D \gets$ GetDependencyMasks($G$) \hfill{\textcolor{blue}{// Find all dependency closures in $G$ and encode them as bitmasks}}\\
$dp \gets [\infty]^{|D|}$,
$prev \gets [-1]^{|D|}$,
$map \gets [\emptyset]^{|D|}$ \\

\For{$i \gets 0$ \KwTo $|D|-1$}{
    \If{$D[i] = \emptyset$}{
        $dp[i] \gets 0$ \\
        \Continue
    }
    \For{$j \gets 0$ \KwTo $i-1$}{
        \If{$D[i] \& D[j] = D[j]$}{
            $stage \gets D[i] - D[j]$ \hfill{\textcolor{blue}{// Extract the set difference of dependencies as a partition}}\\
            $(cost, mp) \gets$ OptimalMapping($stage, R$) \\
            \If{$dp[j] + cost < dp[i]$}{
                $dp[i] \gets dp[j] + cost$ \\
                $prev[i] \gets j$ \\
                $map[i] \gets map[j] \cup mp$ \\
            }
        }
    }
}

$S \gets$ ReconstructSolution($dp, prev, map$) \\
\Return $S$
\caption{DP-based partitioning and mapping}
\label{alg:dp}
\end{algorithm}

\textbf{CG-level Optimization.}
The optimization at this level begins with preprocessing the computation graph through analyzing the operator dependencies within the directed acyclic graph (DAG).
During preprocessing, the compiler first identifies and extracts MVM-based operators, then groups adjacent operators with them to create a condensed CG.
This analysis produces a dependency-preserving linear sequence of operators that forms the foundation for subsequent optimization stages.

To address the capacity limitation inherent in digital CIM architectures, CIMFlow implements a systematic partitioning strategy to divide the model into multiple execution stages.
As detailed in Alg.~\ref{alg:dp}, the model partitioning phase employs a dynamic programming (DP) based approach that optimizes workload distribution across available cores. 
The algorithm incorporates a state compression optimization that encodes all the dependency closures in the DAG as bitmasks, significantly reducing both space complexity and computational overhead.
Each dependency closure represents a self-contained set of operators whose dependencies are fully enclosed within the set, serving as basic building blocks for candidate partitions.

The compiler derives candidate partitions through set operations on these dependency closures, and performs core mapping optimization for each partition.
This process involves strategically duplicating operator weights across clusters of cores when deemed beneficial by the cost estimation model.
To balance parallel execution benefits against communication costs, the estimation model accounts for both computation costs and data transfer overheads across inter- and intra-cluster communications.
These cost assessments and their corresponding optimal mapping configurations are then used to guide the DP-based partition selection.

The final phase of CG-level optimization focuses on inter-core scheduling and intermediate representation (IR) generation. 
The scheduler orchestrates data movement through the NoC interconnection, facilitating the inter-operator pipelines across different clusters.
For each core, the compiler generates an optimized operation sequence incorporating both partitioning decisions and mapping destinations, establishing the foundation for OP-level optimizations.

\textbf{OP-level Optimization.}
Following CG-level workload distribution, the compiler performs fine-grained operator transformations to maximize hardware efficiency. 
This process involves a structured approach that first establishes an ideal mapping in a constraint-free virtual space, and then adapts this mapping to actual hardware resource constraints.

The virtual mapping phase begins by analyzing the dimensional structure of each operator, transforming complex nested loops into a simplified version that aligns with the CIM array structure.
This transformation process maps the software-level weight dimensions onto a two-dimensional array representation.
By temporarily abstracting away physical constraints, the compiler explores the optimal weight data layout strategies, including the image-to-column (\texttt{im2col}) transformation commonly employed in DNN acceleration.

The physical mapping phase then adapts the idealized representation to actual hardware constraints through a series of optimization passes implemented within the MLIR infrastructure. 
The compiler first applies loop tiling based on resource capacity constraints, then systematically extracts MVM operations from the tiled loops for translation into CIM operations.
Through automated analysis, it determines the optimal tile sizes and loop ordering to maximize computational efficiency while respecting resource limitations at each memory hierarchy. 
Memory access operations are then strategically annotated at appropriate loop levels to minimize data transfer overhead.

In the final code generation phase, the optimized IR undergoes conventional compilation techniques, including constant propagation, dead code elimination, and register allocation.
The generated instructions adhere to the CIMFlow ISA specification while realizing the optimized resource mapping decisions, ensuring efficient utilization of the CIM hardware resources.

\subsection{Simulator Design}\label{sec:simulator}
The CIMFlow simulator provides cycle-accurate performance analysis through detailed modeling of the digital CIM architecture across multiple abstraction levels, from individual core execution to chip-level coordination.
Implemented in SystemC~\cite{ieee1666-2023}, the simulator features a detailed pipeline model to track execution flow and resource utilization within each processing unit, while managing parallel execution across cores connected via NoC.
The simulator supports diverse architectural configurations through a user-defined configuration file that adheres to the ISA specifications, while its modular design and standardized interfaces allow straightforward integration of custom architectural components.

At core level, instruction execution follows a three-stage pipeline comprising instruction fetch (IF), decode (DE), and execute (EX).
The EX stage implements detailed execution models for different compute units, each with fine-grained pipelining to enable instruction-level parallelism.
Instruction conflicts and resource utilization are efficiently tracked through a bitmap-based scoring board within the instruction scheduler, ensuring accurate modeling of both computation and data movement patterns.
Through this detailed modeling of both computation and data movement patterns, the simulator provides a comprehensive performance analysis across different architectural levels, tracking metrics such as energy consumption, execution latency, and hardware utilization for each unit.
These detailed insights enable both quantitative evaluation of different CIM design choices and validation of compiler optimizations.
\section{Experimental Results}

\subsection{Experimental Setup}\label{sec:exp-setup}

\begin{table}[t]
\caption{Architecture Parameters of the Default Architecture.}
\label{tab:config}
\renewcommand\arraystretch{1.15}
\resizebox{\linewidth}{!}{
\begin{tabular}{ll|ll|ll}
\hline
\multicolumn{2}{c|}{\textbf{Chip Level}} & \multicolumn{2}{c|}{\textbf{Core Level}} & \multicolumn{2}{c}{\textbf{Unit Level}}   \\ \hline
Core num.          & 64         & CIM comp. unit           & 16 \# MG           & Macro & 512$\times$64          \\
NoC flit size      & 8 Byte     & Macro group           & 8 \# macro          & Element   & 32$\times$8 \\
Global mem.      & 16 MB      & Local mem.      & 512 KB      &         &                         \\ \hline
\end{tabular}
}
\end{table}

To demonstrate how CIMFlow facilitates digital CIM architecture design, we conduct detailed analyses of compilation optimization strategies and present case studies exploring the impact of various architectural and software design choices.
The default architecture parameters are carefully selected to efficiently support typical DNN workload characteristics while maintaining practical hardware constraints, as detailed in Tab.~\ref{tab:config}.
The performance statistics are acquired from multiple industry-standard tools to ensure accurate modeling of all components within the architecture. 
The CIM macro specifications are derived from post-layout analysis based on the design presented in \cite{yan20221}, while other on-chip memory components are evaluated using memory compilers. 
The remaining digital logic modules are implemented in Verilog HDL and synthesized using Design Compiler, with power analysis conducted through PrimeTime PX. 
The NoC interconnection costs are modeled using Noxim~\cite{catania2015noxim}.

We select representative DNN models that span different architectural characteristics and computational demands as our evaluation benchmark. 
The suite encompasses compute-intensive architectures including ResNet18 and VGG19, alongside compact models featuring depth-wise separable convolutions such as MobileNetV2 and EfficientNetB0.
To align with digital CIM implementation constraints, the weights and activations of all models are quantized to INT8.

\subsection{Compilation Optimization Evaluation}

\begin{figure}[t]
    \centering
    \includegraphics[width=\linewidth]{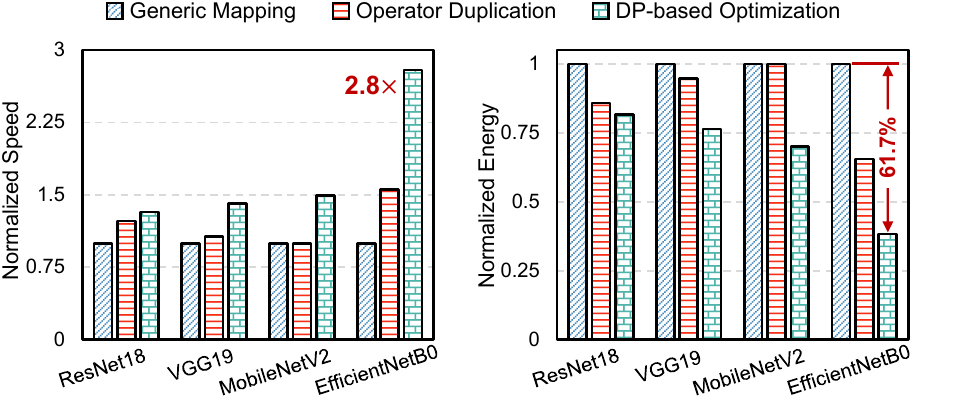}
    \caption{Normalized speed and energy comparison of different compilation optimization strategies across DNN models.}
    \label{fig:exp-comp}
\end{figure}

We evaluate our proposed compilation optimization strategies against two baseline approaches: (1) a generic mapping scheme that implements inter-layer pipeline without operator duplication, and (2) the CG-level partition and opportunistic operator duplication technique from CIM-MLC~\cite{qu2024cim}, which first partitions CG to fit the limited capacity then attempts to utilize vacant resources through weight duplication.
All evaluations use the default architecture configuration described in Tab.~\ref{tab:config} to isolate the impact of compilation strategies.

As illustrated in Fig.~\ref{fig:exp-comp}, our DP-based partitioning and optimization method demonstrates significant performance improvements, achieving up to 2.8$\times$ speedup and 61.7\% energy reduction compared to the baseline approaches.
The benefits are particularly pronounced for compact models like MobileNetV2 and EfficientNetB0, where the conventional partition method proves less effective due to their smaller weight footprints, which leaves fewer unoccupied cores within such partitioned execution stages for duplication opportunities.
This highlights the effectiveness of our DP-based approach in finding optimal partitioning and mapping schemes that maximize performance while respecting SRAM capacity constraints.

\subsection{Architectural Configuration Exploration}

\begin{figure}[t]
    \centering
    \includegraphics[width=\linewidth]{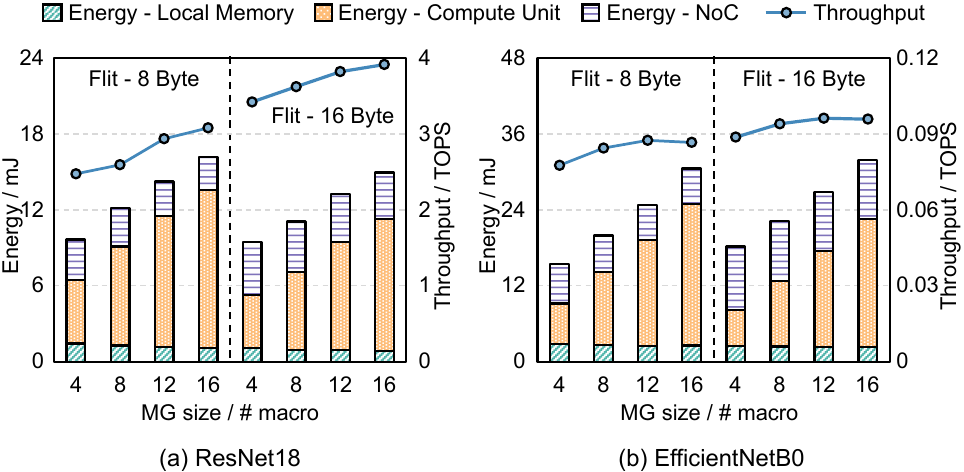}
    \caption{Energy consumption breakdown and throughput across architectures with different MG sizes and NoC link bandwidth.}
    \label{fig:exp-arch}
\end{figure}

The efficiency of digital CIM architectures hinges on the careful balance between computational and data movement capabilities. 
We explore this trade-off through two critical design parameters: MG size scaling and NoC link bandwidth (flit size per cycle) configuration.
Fig.~\ref{fig:exp-arch} presents the energy breakdown and throughput analysis across architectural configurations for both compute-intensive and compact models compiled with the generic mapping strategy.

For ResNet18, increasing MG size consistently improves throughput at the cost of moderately higher energy consumption, with compute unit energy remaining its dominant component.
Doubling the communication bandwidth also boosts inter-layer pipeline throughput by up to 39.6\%.
In contrast, EfficientNetB0 shows different scaling characteristics.
Its lower resource requirements mean that increasing the number of macros per group yields only modest throughput gains, while higher NoC bandwidth introduces substantial data transfer overhead without commensurate performance benefits, consuming up to 55.4\% of total energy consumption when MG size is 4.
These distinct scaling behaviors across models highlight the importance of early-stage architectural exploration, demonstrating the value of CIMFlow as a systematic design and evaluation framework.

\begin{figure}[t]
    \centering
    \includegraphics[width=\linewidth]{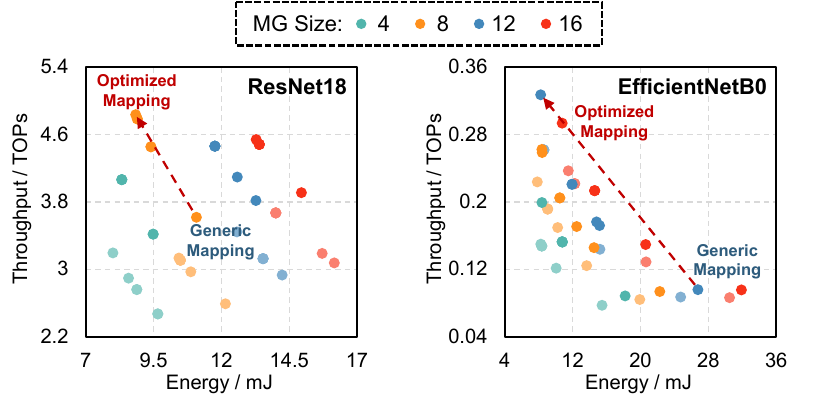}
    \caption{Software/Hardware design space categorized by MG size. Light/dark shades indicate 8/16-byte NoC flit sizes.}
    \label{fig:exp-space}
\end{figure}

To provide insights into the interaction between software and hardware design choices, we further compare different compilation strategies across these hardware configurations.
As shown in Fig.~\ref{fig:exp-space}, while hardware configurations significantly impact the achievable performance envelope, the performance differences between hardware configurations can be significantly reduced or even reversed through careful compilation optimizations.
These observations highlight why an integrated hardware-software co-design approach is essential for digital CIM architectures, as isolated exploration of either space would overlook crucial optimization opportunities.

\section{Conclusion and Future Work}

This paper presents CIMFlow, an integrated framework that enables systematic design and evaluation of digital CIM architectures. 
Through a flexible multi-level ISA design and advanced compilation strategies, CIMFlow effectively addresses key challenges in digital CIM implementation, particularly the SRAM capacity limitations. 
Our experimental results demonstrate the capabilities of our proposed compilation optimizations, achieving up to $2.8\times$ speedup and 61.7\% energy reduction.
The highly customizable and extensible nature of CIMFlow enables systematic design space exploration, providing crucial insights for the software and hardware design of digital CIM.
In the future, we will keep on expanding the framework to support emerging DNN operators and developing automated design space exploration techniques.
We believe CIMFlow represents a significant step toward making digital CIM a practical solution for next-generation AI accelerators.

\clearpage

{
\bibliographystyle{IEEEtran}
\bibliography{ref}
}

\end{document}